\documentclass[%
 reprint,
 amsmath,amssymb,
 aps,
twocolumn]{revtex4-1}

\usepackage{graphicx}
\usepackage{dcolumn}
\usepackage{bm}
\usepackage{amsmath}
\usepackage{xcolor}
\usepackage{physics}

\begin{document}

\preprint{APS/123-QED}

\title{$\Gamma$-Valley Transition-Metal-Dichalcogenide Moir\'e Bands}

\author{Mattia Angeli}
\email{mangeli@sissa.it}
\affiliation{International School for
  Advanced Studies (SISSA), Via Bonomea
  265, I-34136 Trieste, Italy} 
\author{A.H. MacDonald}
\email{macd@physics.utexas.edu }
\affiliation{Physics Department, University of Texas at Austin, Austin TX 78712}

\date{\today}

\begin{abstract}
The valence band maxima of most group-VI transition metal dichalcogenide thin films remain at the
$\Gamma$-point all the way from bulk to bilayer.  
In this paper we develop a continuum theory of the moir\'e minibands that are formed in the valence bands of  
$\Gamma$-valley homobilayers by a small relative twist.  Our effective theory is 
benchmarked against large-scale {\it ab initio} electronic structure calculations that account for lattice relaxation.
As a consequence of an emergent $D_6$ symmetry we find that low-energy $\Gamma$-valley moir\'e holes differ qualitatively from their $K$-valley counterparts addressed previously; in energetic order the first three bands 
realize i) a single-orbital model on a honeycomb lattice, ii) a two-orbital model 
on a honeycomb lattice, and iii) a single-orbital model on a kagome lattice.  
\end{abstract}

\maketitle


{\it Introduction}--- Moir\'e superlattices form in van der Waals heterostructures with twists or lattice constant differences.
In these systems, the moir\'e pattern acts as a long-wavelength modulating potential that alters 
electronic, vibrational and structural properties.
In semiconductors or semimetals the emergent electronic states can be accurately described using continuum model 
Hamiltonians \cite{MacDonald_PNAS} in which commensurability between the moir\'e pattern and the atomic lattice plays no role.
Interest in moir\'e superlattices has increased following the plethora of interesting phenomena \cite{Herrero_1,Herrero_2,Kim_Nature,Wang_arxiv}
discovered in graphene multilayers.
In this Letter we focus on group-VI transition metal dichalcogenides (TMDs) which are currently under active investigation\cite{Dean_TMDS,Li_Nature,Wang_Nature,Yue_Science,Xu_Nature,LeRoy_Nature,Zhu_arXiv,reconstruction}.
In bulk TMDs with 2H structure, the valence band maximum (VBM) is located at the Brillouin-zone center $\Gamma$ point \cite{Yazyev_review}. 
This property is a consequence of the valence band orbital character \cite{Kaxiras_PRB1}, which is dominated by metal $d_{z^2}$/chalcogen $p_z$ antibonding orbitals
whose out-of-plane orientation generates strong inter-layer hybridization that pushes band energies near the $\Gamma$-point up.
The valence band maximum is at the two-dimensional $\Gamma$-point all the way from bulk to bilayer in WS$_2$, MoS$_2$ and MoSe$_2$, the materials on which we focus.
In these cases, the bilayer valence band maxima is a layer-antibonding state 
that is energetically separated with respect to its bonding counterpart by several hundreds of 
meV \cite{supplementary_material}.  This observation motivates the construction of a one-band 
continuum model in which the antibonding state is not explicitly included, and which is 
similar at first sight to the one band Hamiltonian of TMDs heterobilayers \cite{Wu_Hubbard}.
We find however that the emergent symmetries are different in the two cases, and that 
$\Gamma$-valley homobilayers simulate 2D honeycomb lattice physics, opening up a new chapter of 
strong correlation physics in moir\'e superlattices.

\begin{figure}
\includegraphics[width=0.85\linewidth]{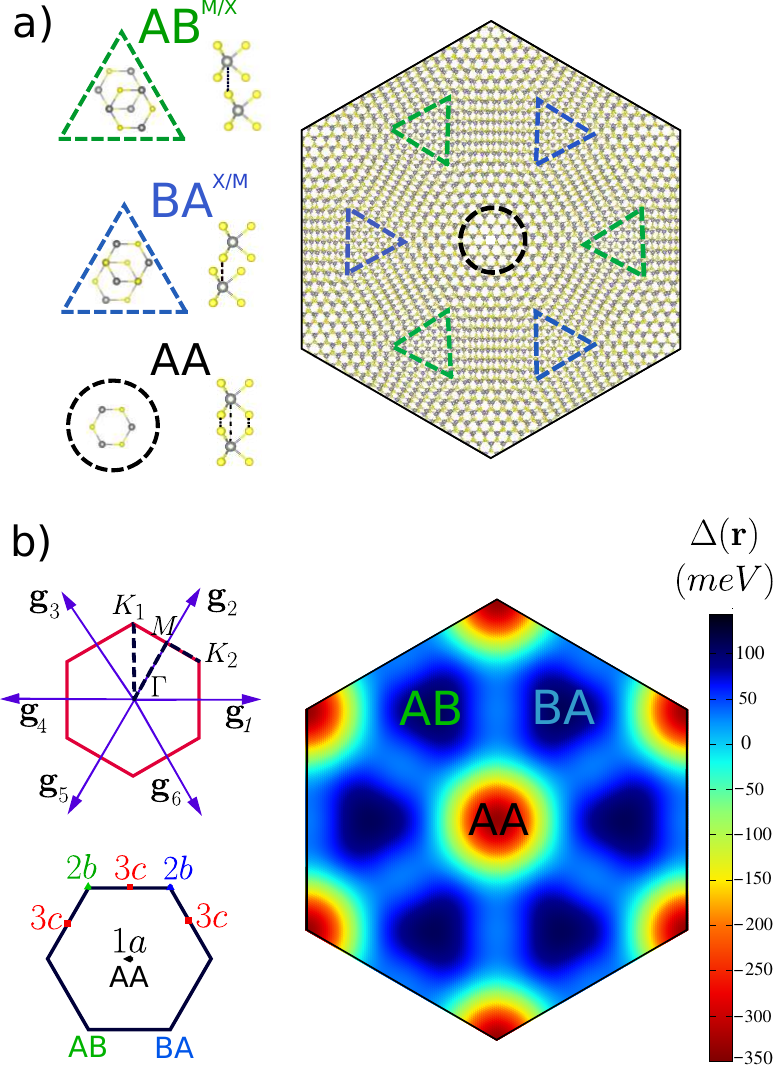}
\caption{a) The three high symmetry configurations in $\beta$ homobilayers: $BA^{X/M}$, $AB^{M/X}$ and AA. 
On the left, the moir\'e pattern formed in a $\theta=3.15^\circ$ homobilayer is illustrated. 
The Bernal stacked regions, whose centers form an honeycomb lattice are denoted by green and blue triangles, and the AA region by a black circle.
b) The first shell ($s=1$) of moir\'e reciprocal lattice vectors used to expand the moir\'e potential and the the maximal Wyckoff positions of wallpaper group 17. 
On the left we show the moir\'e potential $\Delta(\textbf{r})$ for $\text{MoS}_2$, which is attractive for holes 
on the hexagonal network formed by the AB/BA regions.}
\label{superlattice}
\end{figure}

Twisted heterostructures between binary van der Waals monolayers, like TMDs but unlike graphene,
occur in two distinct configurations - referred to here as $\alpha$ and $\beta$ \cite{Rubio_thBN}. 
The two configurations differ by a $180^\circ$ rotation of the top layer with respect to the metal axis. 
In \ref{superlattice}(a) we show a $\beta$ twisted bilayer in which AA regions form a triangular lattice and are 
surrounded by six Bernal ($AB^{M/X}$ and $BA^{X/M}$) regions that form a honeycomb network. 
In the $AB^{M/X}$/$BA^{X/M}$ areas a metal atom (M) on one layer is directly on top/below a chalcogen atom (X) on the other layer.
The two regions are related by a reflection that exchanges the two layers.  In $\alpha$ bilayers, on the other hand,
the ($AB^{M/M}$ and $BA^{X/X}$) Bernal stacked regions are structurally and energetically different. 
In the following we will focus on $\beta$ bilayers and we will omit the apex in the AB/BA labeling.\\

\begin{figure*}
\centering
\includegraphics[width=1\linewidth]{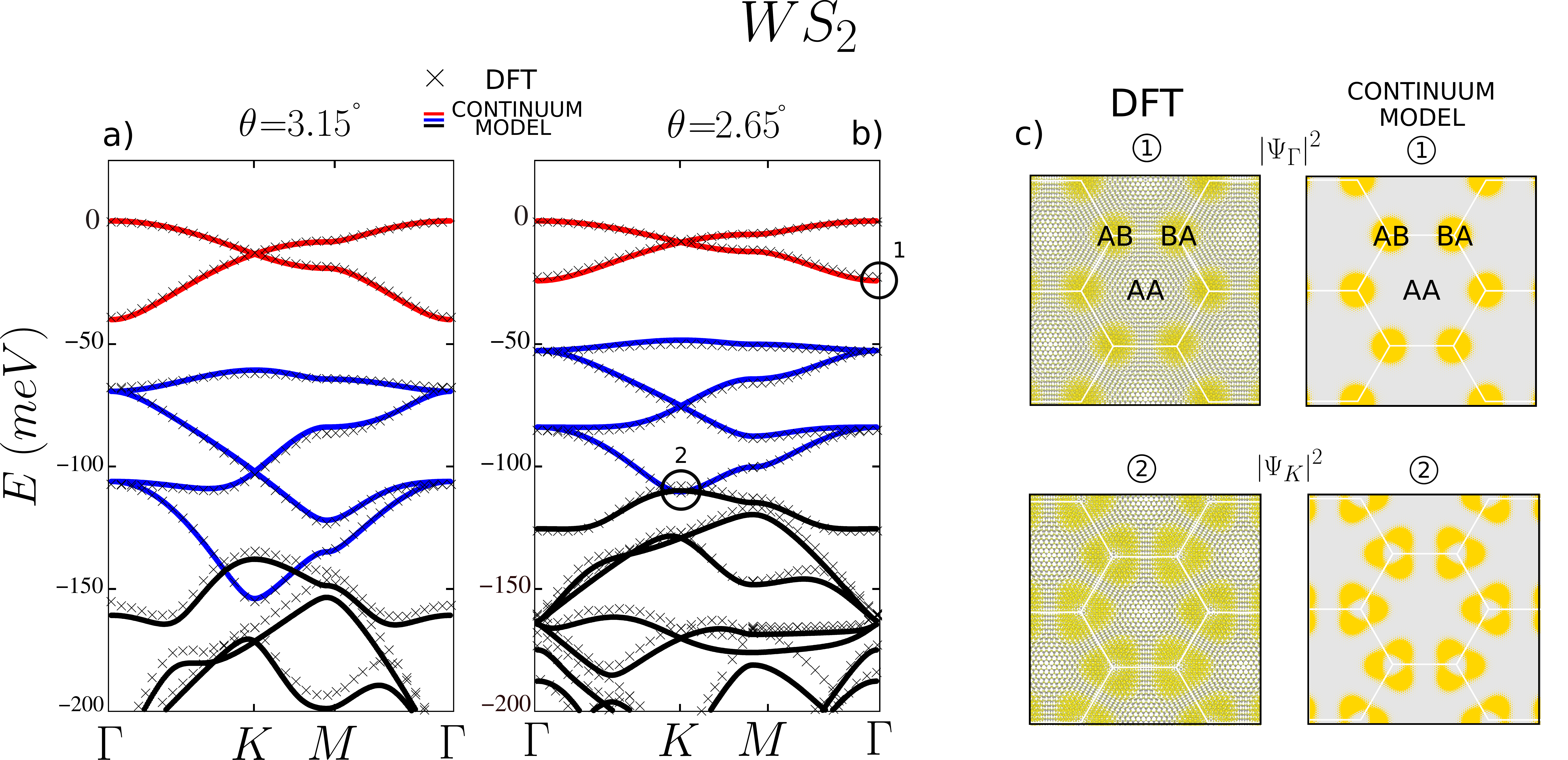}
\caption{DFT vs continuum model.\;
a,b) Twisted $\text{WS}_2$ bandstructure obtained with DFT (black crosses) and continuum model (colored lines) at two different twisting angles $\theta=3.15^\circ,2.65^\circ$. 
Similar plots for twisted $\text{MoS}_2$ and $\text{MoSe}_2$ can be found in Ref.~\cite{supplementary_material}. 
The valence band maximum is set to  $E=0$ and the bands originating from $s$ and $p_x\pm i p_y$ moir\'e Wannier orbitals
are colored in red and blue respectively. c) Charge density distributions of two selected Bloch states encircled in b) 
as obtained with DFT and with the continuum model.}
\label{bands_dftcm}
\end{figure*}

{\it Moir\'e potential symmetries}---
We derive the valence band moiré Hamiltonian from first principles following the approach outlined in \cite{Jung_PRB}. The main difference compared to
the procedure adopted in previous works \cite{Wu_Hubbard,Wu_PRL_2019} is that we obtain continuum model parameters directly from the {\it ab initio} electronic 
structure of fully relaxed twisted bilayers.
In our low-energy model we retain only the anti-bonding state at $\Gamma$, 
which is energetically isolated from other bands by 350-800 meV \cite{supplementary_material} because of interlayer hybridization.  
Neglecting spin-orbit coupling, which vanishes at the $\Gamma$-point by Kramer's theorem, 
we obtain the following simple single band $k \cdot p$ Hamiltonian:
\begin{equation}
  \mathcal{H}= -\frac{\hbar^2 k^2}{2m^*} + \Delta(\textbf{d}),
   \label{Hamiltonian} 
\end{equation}
where $m^*$ is the effective mass and $\Delta(\textbf{d})$ is the potential felt by holes at the valence band maximum 
as a function of the relative displacement $\textbf{d}$ between the two aligned layers.
The two-dimensional lattice periodicity of the aligned bilayers implies that $\Delta(\textbf{d})$ is a periodic function. 
Threefold rotations with respect to the z-axis ($C_{3z}$) require that $\Delta(\textbf{d})$ is equal to $\Delta(C_{3z}\textbf{d})$. 
Moreover, two bilayers stacked by $\textbf{d}$ and $C_{2z}\textbf{d}=-\textbf{d}$
are mapped into each other by a $z\leftrightarrow-z$ mirror and hence have the same bandstructure. This property, which is peculiar to $\beta$ homobilayers, 
further implies that $\Delta(\textbf{d})=\Delta(C_{2z}\textbf{d})$, {\it i.e.} that $\Delta(\textbf{d})$ is a six-fold symmetric function.
As a consequence the moir\'e potential, and hence the Hamiltonian in \eqref{Hamiltonian}, are $D_6$ symmetric objects. 
The extrema of this potential are either at $\textbf{d}=0$ (AA stacking) or at $\textbf{d}=\pm (\textbf{a}_1+\textbf{a}_2)/3$ (AB and BA stacking), 
where $\textbf{a}_1= a_0 (1,0)$ and $\textbf{a}_2=a_0(-1/2,\sqrt{3}/2)$ are the primitive lattice vectors of the monolayer.

Twisting by a small angle $\theta$ yields a local interlayer displacement $\textbf{d} = \theta\hat{z}\times \textbf{r}$.
Replacing $\textbf{d}$ in Eq.~\ref{Hamiltonian} with $\textbf{r}$ then retains the potential's symmetries and 
magnifies positions to yield a moir\'e potential described by the following Fourier expansion:
\begin{equation}
    \Delta(\textbf{r})=\sum_{s} \sum_{j=1}^6 V_s \exp(i \textbf{g}_j^s\cdot\textbf{r} + \phi_s)
    \label{moirè_pot}
\end{equation}
where $\textbf{g}_{j+1}^s=C_{6z} \textbf{g}_j^s$ is the $s$-th shell of six moir\'e $\textbf{g}$ vectors ordered with increasing $|\textbf{g}|$. The phase factors $\phi_s$ are constrained by the $C_{6z}$ symmetry to be  either 0 or $\pi$.
We solve for the moir\'e Hamiltonian Bloch states by expanding in plane waves:
\begin{equation}
    \bra{\textbf{k}+\textbf{g'}}\mathcal{H}\ket{\textbf{k}+\textbf{g}}=-\delta_{\textbf{g},\textbf{g'}}\frac{\hbar^2|\textbf{k+g}|^2}{2m^*} + \Delta(\textbf{g}-\textbf{g}'),
    \label{moirè_H}
\end{equation}
where $\textbf{k}$ is a wavevector in the moir\'e Brillouin-zone.
The applicability of this low-energy model does not rely on commensurability between the moir\'e pattern and the underlying lattice.  
Even though the twisted lattice has only $D_3$ symmetry \cite{Wu_PRL_2019}, the moir\'e Hamiltonian \eqref{moirè_H} inherits the $D_6$ symmetry of the moir\'e potential \eqref{moirè_pot}.
This emergent low-energy property has profound consequences for the low energy moir\'e bands, and is the main focus of this paper.  
The emergence of symmetries not present in the underlying lattice
is a common \cite{Fabrizio_PRB,Senthil_PRB} and intriguing feature of moir\'e materials.\\

\begin{table}[]
    \centering
\begin{tabular}{c|c|c|c}
      & $WS_2$ & $MoS_2$ & $MoSe_2$  \\ 
      \hline
$V_1$ & 33.5 & 39.45 & 36.8 \\
$V_2$ & 4.0 & 6.5 & 8.4 \\ 
$V_3$ & 5.5 &  10.0 & 10.2 \\
$\phi_{1,2,3}$ & $\pi$ & $\pi$ & $\pi$ \\
$m^*$  & 0.87 &  0.9 & 1.17 \\
$a_0$  & 3.18 & 3.182 & 3.295 \\
\end{tabular}
    \caption{Parameters of the moir\'e Hamiltonian (Eq.~\eqref{moirè_H})
    for the three TMD $\beta$-homobilayers considered in this paper. $V_{1,2,3}$ are in meV, $m^*$ is in bare electron mass units,
    and the triangular lattice constant $a_0$ is in Angstroms.}
    \label{table_1}
\end{table}

{\it Ab-inito energy bands}---  
We have performed large scale DFT calculations on the twisted bilayers (see Fig.\ref{bands_dftcm} and \cite{supplementary_material})  using QUANTUM ESPRESSO \cite{QE1,QE2}.
The VBM  at $\Gamma$ is sensitive to the \textbf{d}-dependence of the layer separation, which is substantial\cite{reconstruction,supplementary_material,Falko_PRL,Kaxiras_PRB2,Naik_ArXiv,Naik_PRL,Rubio_tMoS2},
motivating the inclusion of lattice relaxation. Our calculations were performed using the generalized gradient approximation (GGA) with   
the weak van-der-Walls (vdW) forces acting between the layers taken into account by means of the non-local vdW functional vdW-DF2-C09 \cite{Berland_2015,Hamada_PRB} 
(more details on the $ab\;initio$ calculations are given in \cite{supplementary_material}).  The parameters ($V_s$,$\phi_s$,$m^*$) of the continuum model were adjusted to match the DFT bandstructures. 
As shown in Fig.\ref{bands_dftcm} and \cite{supplementary_material}, we found that expanding up to the third shell $s=3$ 
was sufficient to accurately fit the low energy bandstructures
and the charge density distribution of the relaxed bilayers.  The model parameters for twisted $WS_2$, $MoS_2$ and $MoSe_2$ are listed in \ref{table_1}.
As a consequence of the out-of-plane nature of the $p_z$ and $d_{z^2}$ orbitals involved at the $\Gamma$ VBM
the amplitude of the moir\'e potential $V_s$ is more than five times larger than for K-valley TMDs \cite{Wu_Hubbard,Wu_PRL_2019}.
Furthermore, since the VBM in the Bernal stacked regions is higher in energy than in the AA regions (fixing $\phi_s=\pi$), 
the minimum of the potential felt by holes in the valence bands lies on the hexagonal lattice formed by the AB/BA regions (see Fig.\ref{superlattice}(b)). 
The physics of the moir\'e band edges in $\Gamma$-valley $\beta$ TMDs homobilayers is therefore generated by orbitals sitting on a 
honeycomb lattice.\\

\begin{figure}
\includegraphics[width=0.95\linewidth]{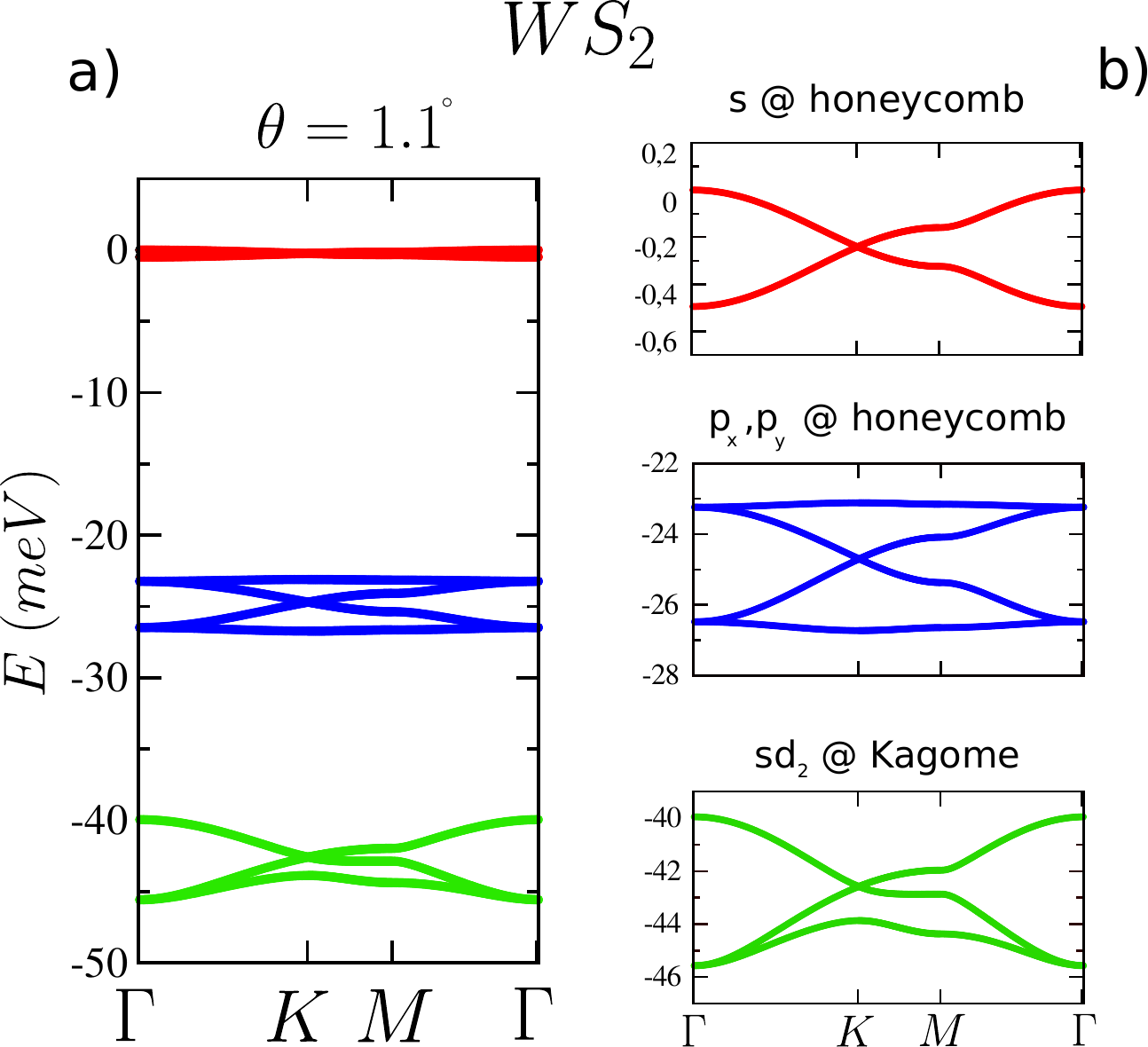}
\caption{a) Bandstructure of $\theta=1.1^\circ$ twisted $WS_2$. b) The first set of bands has a bandwidth of $\approx 0.5\; meV$ and is formed by $s$ orbitals on an honeycomb lattice.  The second is formed by $p_x\pm i p_y$ orbitals on the honeycomb and  has a bandwidth of $\approx 3.2\; meV$ . The third set of bands, formed by a set of hybridized $sd_2$ orbitals on a kagome lattice, has a bandwidth of $\approx 5.75 \;meV $.
}
\label{bands_continuum}
\end{figure}

{\it Twist-angle dependence}--- We now demonstrate the efficacy of the continuum model by 
using it to predict the bandstructure of twisted TMDs at angles where full microscopic calculations are prohibitive.
To reveal the moir\'e band physics more fully we employ Topological Quantum Chemistry theory \cite{Bernevig_Nature} to identify the symmetries and centers of 
the Wannier orbitals underlying the moir\'e bands by i) computing the symmetry of the Bloch states
and classifying them in terms of the irreducible representations (irreps) of the little groups at the corresponding high symmetry points
and ii) comparing the list of irreps with the Elementary Band Representatons (EBR) of the space group $P6mm$ listed on the 
Bilbao Crystallographic server \cite{Bilbao_server}.
In Fig.\ref{bands_continuum} the band structure of $\theta=1.1^\circ$  twisted $\text{WS}_2$ is shown.
Consistent with the emergent honeycomb structure of the moir\'e potential, the first set of bands is formed by a pair of s-like orbitals centered in the AB/BA regions. 
These bands are non-degenerate at $\Gamma$, form a Dirac node at K, and are topologically equivalent to the $\pi$ bands of graphene.
Consistent with previous studies on $\text{MoS}_2$ \cite{Rubio_tMoS2}, the second set of bands, is instead formed by $p_x\pm i p_y$ orbitals on an honeycomb \cite{WU1}
that form a pair of almost dispersionless bands and also have a Dirac node at K.
The third set of bands is even more intriguing, because it is formed by an odd (three) number of bands, a feature inconsistent with orbitals on an honeycomb. 
The symmetry analysis reveals that they are generated by orbitals centered on the $3c$ Wyckoff positions, which lie at the mid point between two honeycomb sites (see Fig.\ref{superlattice} b). 
Interestingly, the lattice formed by this Wyckoff positions is a kagome lattice, a prominent platform to host frustration and spin liquid physics \cite{SpinLiquid,Quantum_spin_liquids}. 
The topology of these bands, which have one flat band and a Dirac node, further confirms the kagome picture.
To understand these bands in terms of the hexagonal moir\'e potential of the system, we need two orbitals on different honeycomb sites to hybridize. 
As a consequence of this hybridization, the Wannier centers move from the honeycomb AB/BA sites (2b) to their midpoints (3c), effectively turning the honeycomb into a kagome lattice.

\begin{figure}
\includegraphics[width=1.05\linewidth]{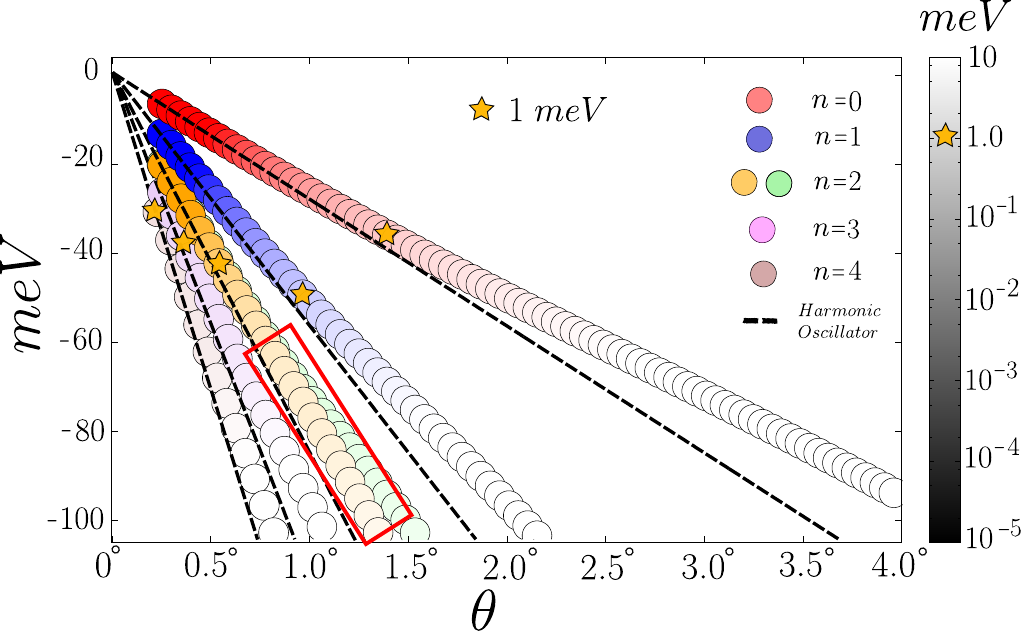}
\caption{Twisted TMDs as coupled harmonic oscillators: Band center of the five highest in energy valence bands in twisted $MoS_2$ as a function of twist angle. 
The bandwidth is encoded in color with the 1 meV bandwidth threshold marked by a star. The black dashed lines correspond to the $n=0-4$ harmonic oscillator eigenvalues 
with an oscillator frequency that is proportional to twist angle. Within the red rectangle, the $n=2$ levels
split in two sets of kagome bands (in green and orange) as a consequence of the $sd_2$ orbital hybridization.}
\label{HOSC}
\end{figure}

We have also performed a systematic analysis of all the bands within $\approx 110$ meV below the VBM as a function of twist angle.
This analysis is summarized in Fig.\ref{HOSC}, where the band centers of the first five sets of bands are represented
by colored circles whose saturation represents the corresponding bandwidths.  The band centers and separations 
asymptotically evolve linearly with the inverse moir\'e
length $a_M^{-1} \propto \theta$, whereas the bandwidths decrease exponentially with $\theta$.
This behaviour can be understood by making an harmonic approximation to the moir\'e potential 
near its maxima \cite{Wu_Hubbard,Carr_arxiv}; 
$\Delta(\textbf{r}) \approx -\gamma (\textbf{r}/a_M)^2/2$ where $\gamma=8\pi^2(V_1+6 V_2+4 V_3)$ for $\beta$-homobilayers.
The black dashed lines in Fig.\ref{HOSC} are the $n=0-4$ eigenvalues of this harmonic oscillator problem.  The good agreement identifies the harmonic oscillators wavefunctions as Wannier functions. 
Within this approximation, the ratio between the Wannier wavefunction size to the moir\'e period, $a_W/a_M \approx \theta^{1/2}$, implying that the overlap between 
neighboring Wannier functions and the bandwidths are $\propto e^{-\theta}$.
The absence of magic angles in this systems is in sharp contrast with twisted bilayer graphene case\cite{Vishvanath_PRL,Kaxiras_PRR,Fabrizio_PRB,Bernevig_PRL,Fabrizio_PRX}.
The symmetries of the 2D harmonic oscillator wavefunctions further confirm this picture since the
$n=0$ orbital is an $s$ state and the $n=1$ doublet is spanned by $p_x \pm i p_y$.
The $n=2$ harmonic oscillator orbitals consist of one $s$ and two $d$ orbitals per honeycomb site.
The EBR analysis shows that over a broad intermediate twist angles ($0.85^\circ  < \theta < 1.4^\circ$) the six $n=2$ orbitals per unit cell
separate into two groups of three which can be identified as $sd_2$ bonding and antibonding bands centered on 
kagome lattice sites located half-way the honeycomb sites.  This situation is known to give rise to kagome lattice physics \cite{Liu_PRL}, but has not been 
realized experimentally.  The splitting between bonding and antibonding bands decreases with the angle until, at a critical $\theta_c \approx 0.85$, the six bands merge.
By further decreasing the angle the bands disconnect again, this time forming sets of 2 and 4 bands separated by a small gap, whose dispersion resemble that of the $s$ and $p_x \pm i p_y$ bands.  
We observed similar behaviour also in the $n=3,4$ sets of energy bands. This feature is due to high order terms not included in our simple harmonic oscillator approximation.
In particular, the topology of  these  sub-bands is always that induced by orbitals ($s$ or $p$) whose angular momenta is lower than 2, 
consistent with the constraints imposed by a triangular quantum well \cite{triangular}. 
We expect this behaviour to be even further enhanced by the strong reconstruction observed at very small angles, 
which tends to expand and sharpen the triangular Bernal domains \cite{reconstruction}.

{\it Discussion}--- We have shown that because of an emergent $D_6$ symmetry, 
twisted $\Gamma$-valley TMDs open up a new frontier in strong-correlation moir\'e superlattice 
physics. The highest in energy moir\'e valence bands provide a convenient realization of artificial graphene, 
but with a fine-structure constant that can be tuned across the chiral symmetry breaking 
phase transition\cite{Chiral1,Chiral2} simply by varying twist angle.  The second set of bands is formed by $p_x \pm i p_y$ orbitals on a honeycomb lattice \cite{WU1} and 
is a promising candidate to study orbital and nematic order \cite{Nematic,Rubio_tMoS2,Venderbos,Natori} . 
Thanks to an $sd_2$ hybridization of honeycomb lattice orbitals, 
the third set of bands realizes a kagome lattice model and is expected to host spin-liquid physics\cite{SpinLiquid, Quantum_spin_liquids}.  
Since all models have bandwidths that can be adjusted simply by varying twist angles or by applying pressure, they provide an 
enticing platform to study the  exotic properties of strongly correlation physics on the honeycomb lattice \cite{WU2,WU3,Gregory,Natori,Kiesel}. 
Furthermore, by inducing spin-orbit coupling with gate electric fields or by proximity, all bands can exhibit topologically 
non-trivial states \cite{Tang,Iimura,Kane,Wang}.
Finally, the Hamiltonian presented here can be easily generalized to describe $\Gamma$ $\alpha$-homobilayers or heterobilayers by  
simply relaxing the ($\phi_s=0,\pi$) constraint imposed by the emergent $D_6$ symmetry.

\begin{acknowledgments}
MA acknowledges funding from SISSA and is extremely thankful to the University of Texas at Austin for the hospitality received. MA acknowledges useful discussions with M. Fabrizio and A. dal Corso.
We acknowledge HPC resources provided by the CINECA Computing center in Bologna (Italy) and the Texas Advanced Computing Center (TACC) at The University of Texas at Austin (USA).
AHM acknowledges support from DOE grant DE- FG02-02ER45958 and Welch Foundation grant TBF1473.

\end{acknowledgments}

\appendix



\section{Supplementary Material}

\subsection{Bandstructure {\it vs.} Stacking}

\begin{figure}[h]
\centering
\includegraphics[width=0.9\linewidth]{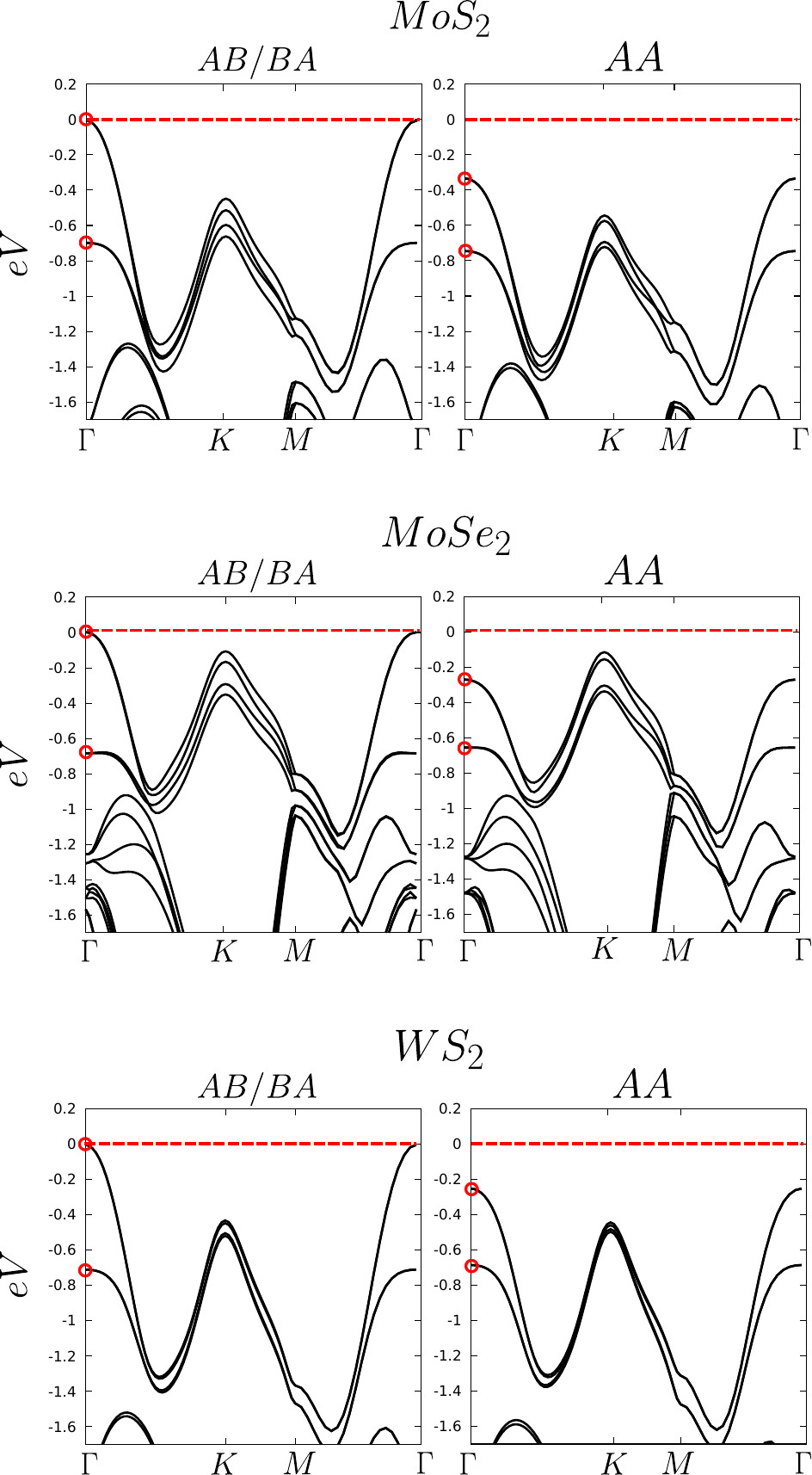}
\caption{Bilayer DFT bandstructures for relaxed $MoS_2$, $MoSe_2$ and $WS_2$ with AA and AB/BA stacking. The VBM energy is denoted by a red dashed line while the layer bonding and antibonding valence states at $\Gamma$ are denoted by red circles.}
\label{figS1}
\end{figure}

We performed a systematic analysis of the stacking dependence of untwisted bilayer electronic structure to identify the 
valence band edges at two high-simmetry stackings (AA and AB/BA). 
The calculations were performed with QUANTUM ESPRESSO \cite{QE1,QE2} using the Perdew-Burke-Ernzerhof (PBE) generalized gradient approximation (GGA)  functional. 
The weak van-der-Walls (vdW) forces acting between the layers were taken into account by means of the non-local vdW functional vdW-DF2-C09 \cite{Berland_2015,Hamada_PRB}, which has been shown to give good results over a broad range of layered materials \cite{Marzari,Cantele}. 
Spin orbit coupling (SOC) is included in these calculations. For all the elements considered except Tungsten (W)
we used fully relativistic ultrasoft pseudopotentials from pslibrary.1.0.0 \cite{DalCorso}; the Tungsten (W) pseudopotential was instead taken from the SG15-ONCV \cite{SG15} library.
The plane-wave basis used to
sample the reciprocal space was selected with a kinetic energy
cutoff of $55\; Ry$ for the wave functions, while a $500\;Ry$
cutoff energy was is used to represent the charge density.
The BZ was sampled by an $20 \times 20 \times 1$ Monkhorst-Pack k-point grid.
We initially relaxed the monolayer unit cells in order to determine the lattice parameters $a_0$. 
Then, we relaxed the bilayers geometry with a vacuum layer of $20\; \AA$.
In Fig.\ref{figS1} we show the bandstructures of AA and AB/BA stacked $MoS_2$,$MoSe_2$ and $WS_2$. 
As can be seen, the VBM is always located at the $\Gamma$ point. 
At this point, the orbital character is dominated by out-of-plane oriented metal $d_z$ and  chalcogen $p_z$ orbitals, so that the VBM is strongly sensitive to
the interlayer distance which is larger by $\approx 0.7 \AA$ in the AA stacking case. This leads to a $300-350 \; meV$ $\Gamma$ energy difference between the two
configurations.  Finally, we emphasize with red circles the large separation between 
layer bonding and antibonding states at $\Gamma$. We checked their energy separation for bilayers with a generic 
relative stacking vector $\textbf{d}$ taken on a 10x10 real space grid of the unit cell vectors.
We found that the lowest energy separations occur for AA stacking, and that the separation is never lower than $350-400 \; meV$, thus justifying our one-band model assumption.

\subsection{DFT calculations on the twisted supercells}

\begin{figure*}[ht]
\centering
\includegraphics[width=0.9\linewidth]{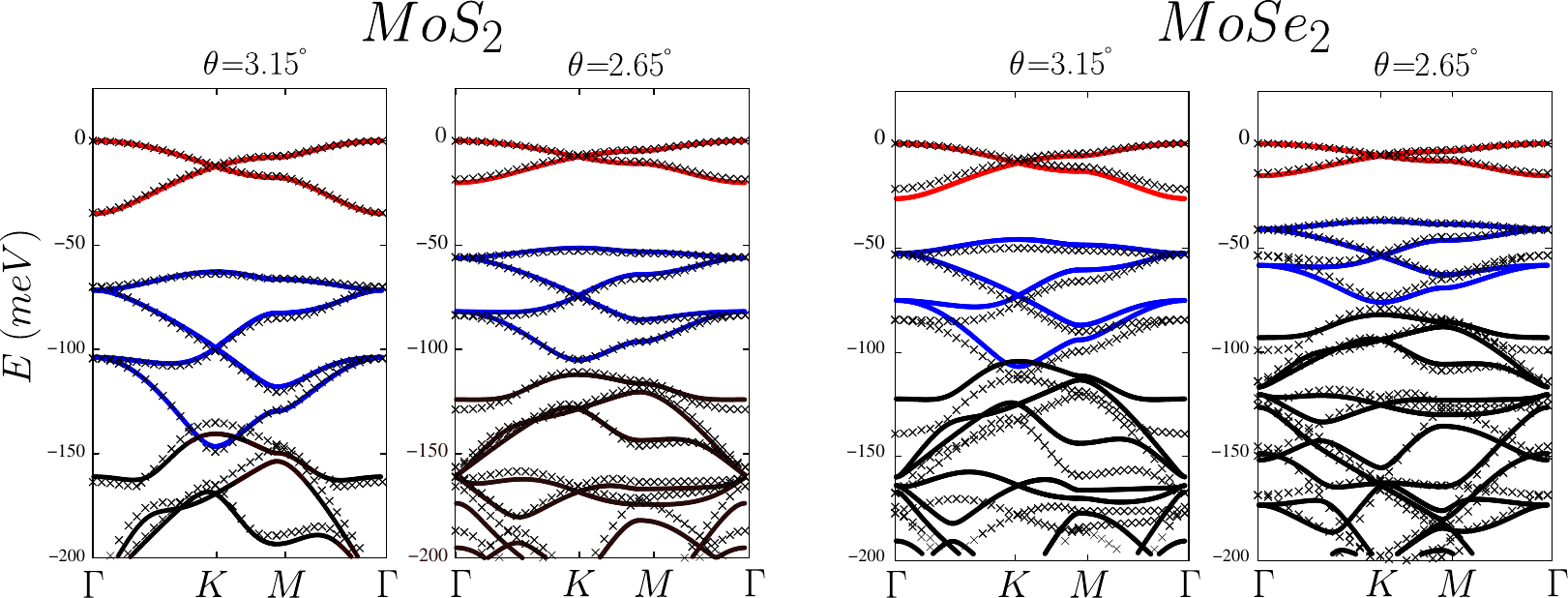}
\caption{ Twisted $MoS_2$  and $MoSe_2$ bandstructures obtained with DFT (black crosses) and continuum model (colored lines) at two different twist angles $\theta=3.15^\circ,2.65^\circ$. 
The valence band maximum is set to  $E=0$ and the bands originating from $s$ and $p_x\pm i p_y$ moir\'e Wannier orbitals
are colored in red and blue respectively}
\label{figS2}
\end{figure*}

We performed large scale ab-initio calculations on the relaxed $\theta=2.65^\circ , 3.15^\circ$ twisted supercells.  Due to the large volume of these supercells, containing more than 2800 atoms, these calculations required massive parallelization over more than 4000-5000 CPUs.
Although most of the details of these calculations are similar to those described in the previous section, here we did not include SOC, whose effect vanishes at $\Gamma$ (see Fig.\ref{figS1}) due to Kramer's theorem.  Instead we used scalar relativistic pseudopotentials from the SSSP efficiency library \cite{SSSP},
which required lower plane wave and charge density kinetic energy cutoffs equal to $40\;Ry$ and $320\; Ry$ respectively.
We sampled the BZ only at the $\Gamma$ point.
The atomic positions were optimized
 with the convergence criterion that all components of the forces acting on the atoms must be less than $10^{-3}$ Ry/bohr. 
 The relaxation was performed along all the spatial directions in order to capture not only the interlayer corrugation but also the local in-plane strain patterns 
 which play an important role in these systems \cite{Naik_ArXiv}.
In Fig.\ref{figS2}  we compare the DFT bandstructure and the continuum model fits for twisted $MoS_2$ and $MoSe_2$.


\bibliographystyle{apsrev}

\end{document}